\documentclass[journal,10pt,letterpaper,final,twocolumn]{IEEEtran}%
\usepackage{amsmath}
\usepackage{amsfonts}
\usepackage{cite}
\usepackage{amssymb}
\usepackage{mathrsfs}
\usepackage{ragged2e}
\usepackage{graphicx}
\usepackage[usenames,dvipsnames]{color}
\usepackage{epstopdf}
\usepackage[all]{xy}
\usepackage[]{mcode}
\usepackage{booktabs}
\usepackage{supertabular}
\usepackage{subcaption}
\usepackage{multirow}
\usepackage{array}

\providecommand{\U}[1]{\protect\rule{.1in}{.1in}}
\pdfoutput=1
\providecommand{\U}[1]{\protect\rule{.1in}{.1in}}

\newcommand{\qed}{\nobreak \ifvmode \relax \else
      \ifdim\lastskip<1.5em \hskip-\lastskip
      \hskip1.5em plus0em minus0.5em \fi \nobreak
      \vrule height0.75em width0.5em depth0.25em\fi}
      
\begin{document}

\title{\LARGE PHY Research Is Sick but Curable: An Empirical Analysis}
\author{Kevin Luo, Shuping Dang, \textit{Member, IEEE}, Basem Shihada, \textit{Senior Member, IEEE}, \\Mohamed-Slim Alouini, \textit{Fellow, IEEE}
  \thanks{K. Luo is with Graduate School of Economics, Kobe University, Kobe 657-8501, Japan (e-mail: kevin.luo@stu.kobe-u.ac.jp).
  
  S. Dang, B. Shihada and M.-S. Alouini are with Computer, Electrical and Mathematical Science and Engineering Division, King Abdullah University of Science and Technology (KAUST), 
Thuwal 23955-6900, Kingdom of Saudi Arabia (e-mail: \{shuping.dang, basem.shihada, slim.alouini\}@kaust.edu.sa).}}

\maketitle

\begin{abstract}
The controversy and argument on the usefulness of the physical layer (PHY) academic research for wireless communications are long-standing since the cellular communication paradigm gets to its maturity. In particular, researchers suspect that the performance improvement in cellular communications is primarily attributable to the increases in telecommunication infrastructure and radio spectrum instead of the PHY academic research, whereas concrete evidence is lacking. To respond to this controversy from an objective perspective, we employ econometric approaches to quantify the contributions of the PHY academic research and other performance determinants. Through empirical analysis and the quantitative evidence obtained, albeit preliminary, we shed light on the following issues: 1) what determines the cross-national differences in cellular network performance; 2) to what extent the PHY academic research and other factors affect cellular network performance; 3) what suggestions we can obtain from the data analysis for the stakeholders of the PHY research. To the best of our knowledge, this article is the first `empirical telecommunication research,' and the first effort to involve econometric methodologies to evaluate the usefulness of the PHY academic research.
\end{abstract}


\section{Introduction}
Telecommunications have experienced a high-speed developing epoch in the past decades, and the constructive influence of this field on human's society and technological development can never be overestimated. Among a variety of taxonomic branches of telecommunications, cellular communications have become ubiquitous and increasingly related to the daily life of everyone. Due to the importance of cellular communications, enormous financial resources and human capital have been devoted into academia for scientific research over the past decades \cite{dth054,6824752}. As a result of these investments, a number of ground-breaking communication technologies were invented and implemented to significantly ease our daily communications nowadays, including multiple-input and multiple-output (MIMO) system, orthogonal frequency-division multiplexing (OFDM), and software-defined networking (SDN). As the cellular concept becomes increasingly ripe, there exists another voice questioning the usefulness of the physical layer (PHY) academic research for future wireless communications \cite{dth054}, in particular, the relation between the PHY research activities and service improvements on practical cellular implementations \cite{5741160}. 

\begin{figure}[!t]
\centering
\includegraphics[width=3.5in]{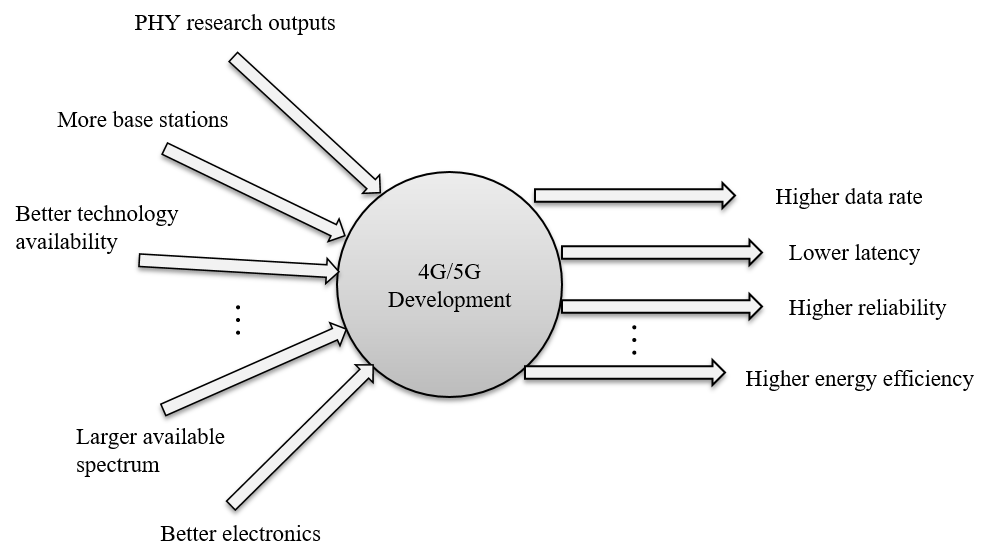}
\caption{Summary of the 4G/5G development from a systematic viewpoint.}
\label{sys}
\end{figure}

Fig \ref{sys} summarizes the determinants of the 4G/5G development from a systematic viewpoint, in which the technology evolution is affected by multiple factors and yields many performance advantages. However, the system \textit{per se} is still a black box to observers, because of the sophisticated and interconnected relation among all factors. Although technological breakthroughs are of great importance, it is conjectured that the dominant factors for improving practical cellular systems in recent years refer to a higher number of base stations (BSs) and a wider radio spectrum in use \cite{4623708,6824752,7980729}. The former is a capital-intensive growth pattern, and the latter is more related to  electronics and manufacturing industry \cite{923566}, which might not be directly associated the PHY research. 

Together with the `5G enthusiasm,' a number of  novel PHY communication conceptions, paradigms, and technologies have been proposed and investigated in depth. Are they realistic or nothing but hypes? There are ample examples that a number of heated technologies in the past have finally turned out to be hyped, and wireless communications are not exceptional \cite{su10103472,4468730}. These research topics and activities cost enormous financial and human resources after all. As a result, the controversy and argument on the usefulness of the PHY academic research are long-standing and constructive \cite{dth054}. In this light, from the research stakeholders' perspective, it is instinctive to concern the following questions:
\begin{itemize}
\item What are the factors determining the cross-national differences in telecommunication performance?
\item To what extent does the PHY academic research affect the cellular network performance?
\item As an engineering discipline, what can we do to maximize the usefulness of the PHY academic research?
\end{itemize}
To provide some clues and preliminary answers to these questions in an objective manner, we employ a set of methodologies of econometrics to carry out analysis. To the best of our knowledge, the study presented in this article is the first `empirical telecommunication research' that quantitatively investigates the deductions of telecommunication  theory; 
it is also the first empirical attempt to evaluate the usefulness of the PHY research and quantify the contributions of different determinants to cellular network performance.

\section{Literature Review and Contributions}
There is an age-old concern of the usefulness of academic research in shaping ripe technological paradigms. Just as the current suspicion of 5G and 6G, researchers in the 3G era have already questioned 4G for whether it has suffered a lot of trendy buzzwords \cite{923566}. Meanwhile, it is noticed in \cite{923566} that the evolution of wireless communications does not fully obey Moore's Law, even though the performance of wireless communication systems is highly related to the processing capability of electronics installed. This is because other non-technological factors have an equal or even more significant impact on the performance of wireless communication systems, including standardization, legislation, intellectual property protection, legacy deployments, and a series of complex socio-economic and human factors. More importantly, the performance of wireless communication systems is strictly constrained by basic physical laws of electromagnetic propagation and electronics. 

In \cite{4468730}, it is further pointed out that only a small portion of wireless technologies have finally become commercially viable civil products and appeared on the market. Overall, timing is critical for successful applications of novel wireless technologies. This article also introduces the roles of academia, industry, and government for wireless technology  implementation. In addition, the authors study the demands, application scenarios, and marketing incentive schemes. More crucial discussions are also provided around the measures for quantifying the usefulness of academic research and funding strategies.

Estimated according to Cooper’s Law in \cite{4623708}, the main contributors to capacity are reduced cell size, wider radio spectrum, and better modulation schemes, among which the reduced cell size dominates. Following the statement breaking down the million-fold capacity increase over the past 60 years in \cite{4623708}, \cite{5741160} presents a critical and detailed retrospect of the progress in 2G, 3G, and 4G. In this article, the life-or-death problem of PHY is investigated. It is stated that the PHY academic research indeed hits the performance wall of the Shannon–Hartley theorem, and any endeavor aiming at increasing capacity would yield marginal returns. The authors suggest that the PHY research should embark onto solving realistic problems, instead of creating academic problems. It has also been stressed that more practical considerations have to be taken into the PHY research, so as to be instructive. Furthermore, it is concluded that the PHY academic research is still alive, but morbid. The authors in \cite{7980729} continues to discuss the practical aspects of the PHY research and wireless communications in the context of 5G and examine the key performance indicators (KPIs) of cellular communications.

On the other hand, it is emphasized in \cite{6824752} that the PHY academic research is of high importance, because we need to understand and model the behavior of high-frequency radio signal as well as to manage cross-cell interference when densifying communication networks in space and frequency spectrum. Also, the crucial role of the PHY research in engineering education and socio-technical milieu of wireless technology cannot be overlooked \cite{8750780}. Although insightful and thought-provoking, the analyses and discussions of the pros and cons regarding the PHY academic research in all aforementioned articles are mainly given in a subjective manner, lacking rigorous analysis and quantitative evidence. 

To obtain quantitative evidence, many researchers have borrowed the analytical frameworks of sociology in studying telecommunication related issues. Employing regression analyses based on empirical data, \cite{gruber2001diffusion} studies the determinants of the diffusion of mobile telecommunication services in the European Union. The authors find that technological breakthroughs, expansion in frequency spectrum, and the increased market competitiveness are essential for the diffusion of 3G mobile services. The empirical data also demonstrates a simultaneous increase of BSs and bandwidth in the 3G era and predicts a robust growth of telecommunication infrastructure in the near future.

To address the effective cooperation between academia and industry, \cite{dth054} conducts an empirical analysis to examine the localized interrelations between academic research and entrepreneurial innovation in the wireless sector. The authors find evidence of regional co-localization of scientific and industrial specializations, as well as the positive feedback between academic publications and patent output. However, publications and patents can be flawed measures of innovation output since they do not necessarily imply useful technologies \cite{acs1992real}. 

The authors in \cite{shin2011socio} employ the sociology of translation to document the socio-technical dynamics of 4G systems in Korea. A set of factors affecting the evolution of 4G are hereby identified from both supply and demand sides. The authors in \cite{Frias2012} present a techno-economic analysis for 4G/LTE networks. In particular, the authors evaluate the economic gains of building different types of BSs by using a techno-economic model and give a flow diagram detailing the techno-economic analytical procedure for 4G/LTE cellular systems. Both studies provide generic but subjective analytical frameworks in evaluating the development of wireless technology.

Empirical methods have enriched the analytical tools in the field of wireless communications. Overall, existing empirical research works regarding wireless communications have two limitations. First, previous empirical attempts are oriented to address only socio-economic issues in the wireless sector (e.g., the determinants of mobile service diffusion), instead of the research topics in telecommunication science. As a result, there is an urgent need to extend the empirical knowledge on communication theories. Second, the empirical evidence, in particular quantitative evidence on the status quo of the PHY research remains scarce, and much of the previous evidence was obtained based on anecdotes and questionable observations (e.g., improper metrics for innovative output). Hence, unified empirical frameworks and solid data sources are indispensable for rigorous empirical assessments of the PHY research status. 

To overcome the above limitations, the study in this article pays careful attention to the consistency between theoretical deduction and empirical modeling, the validity of empirical methods, and metric definitions in quantifying the relative importance of the PHY research and other performance determinants.

\section{Empirical Analysis}
After decades of path-breaking advances, academia begins to sense the diminishing return and even stagnation in wireless communications \cite{5741160}. There is widespread speculation that the impressive improvements in mobile communication over the past decades were primarily achieved by the densification of BSs and the increase in spectrum resource (a capital-intensive growth pattern), instead of the exponential growth in published papers and information and communications technology (ICT) patents (a technology-intensive growth pattern). So far, quantitative evidence is still lacking and no consensus has been achieved yet.

As an applied discipline, the validity of the PHY layer research should be judged by its contribution to the cellular network performance. To evaluate practical scenarios in a unified and simple framework, econometric analysis is a promising option. Econometric analysis is  midway path between simulation and control experiment. It is based on real-world settings instead of parameter assumptions, and it derives parameter estimates in a quasi-experimental manner. We begin with the theoretical modeling of network throughput (the key performance indicator of cellular networks nowadays) and then introduce our estimation method and models based on econometrics.

\subsection{Theoretical and Empirical Modeling}
Stemming from the Shannon–Hartley theorem, the cellular concept was proposed to boost network throughput by limiting the service coverage of each BS to a hexagonal cell, so that scarce frequency spectrum can be reused in different cells with a sufficiently large physical separation. As a result, for a given geographic region, network planners can theoretically provide an arbitrarily high throughput by increasing the number of BSs with a smaller transmit power by a proper frequency reuse scheme \cite{s2010wireless}. Therefore, assuming that only the interference from the first-layer interfering cells is concerned and that all BSs are homogeneous with the same transmit power $P_t$, the downlink normalized network throughput per user over a given geographic region can be approximated as
\begin{equation}\label{djashk23ad111}\small
C\approx\frac{KM}{UN}\log_2\left(1+\frac{(3N)^{\alpha/2}}{6}\right),
\end{equation}
where $K$ is the total number of available channels; $N$ is the number of cells that collectively use these $K$ available channels; $U$ is the number of users subscribing downlink data service; $\alpha$ is the path loss exponent, which characterizes the quality of signal propagation environment; $M$ is the total number of BSs over the entire geographic region required to be covered, and the following restriction shall be satisfied in order to provide appropriate quality of service (QoS): $M>A\rho P_t^{-\alpha/2}$, where $A$ is the area of the given geographic region required to be covered and $\rho$ is a parameter related to the network and system configurations.

Cooper's Law leads to a speculation that the million-fold improvement in data rate over the past 60 years can be primarily attributed to the densification of BSs (which accounts for a 1600-fold gain) and the expansion of radio spectrum (which accounts for a 30-fold gain), whereas the progress in communication theory has contributed no more than a 10-fold gain \cite{4623708}. Consistent with  Cooper's Law, we construct the econometric model of throughput and the corresponding empirical model as
\begin{equation}\label{ajhajhj2dsa2d1a}\small
\begin{cases}
&C_i=\beta_1\frac{K_iM_i}{U_iN_i}+\varepsilon_i\\
&\\
&\mathrm{Speed}_i=\beta_1\mathrm{BaseStation}_i\\
&+\beta_2\mathrm{Spectrum}_i+\beta_3\log\left(\mathrm{Research}_i\right)+\delta_i
\end{cases}
\end{equation}
where $\beta_{(\cdot)}$ is the coefficient estimate, which characterizes the impact of the associated factor on the metric; the subscript $i$ is the country-specific index; $\mathrm{Speed}_i$ denotes the average mobile download speed in Mbps, which is a perceived KPI of network throughput per user; $\mathrm{BaseStation}_i$ is the number of installed telecommunication towers per 1000 mobile subscribers, which represents the level of telecommunication infrastructure; $\mathrm{Spectrum}_i$ is the overall downlink bandwidth  in operation; $\mathrm{Research}_i$ is the cumulative outputs of the PHY research, represented by the numbers of published papers, highly cited papers, or ICT patent applications; $\varepsilon_i$ and $\delta_i$ are the error terms.

Consistent with \cite{6824752}, the models constructed in (\ref{ajhajhj2dsa2d1a}) imply that the increases in the BS density and bandwidth have a proportional payoff in terms of throughput, which leads to the benchmark empirical model of the network throughput.

\subsection{Empirical Analysis and Discussion}

\begin{table*}[!t]
\renewcommand{\arraystretch}{1.3}
\caption{Definitions and sources of the key metrics.}
\label{biaogeyi}
\centering
\begin{tabular}{c|c|c|c|c|c|c}
\hline
Metric & Average & \begin{tabular}{@{}c@{}}Standard \\ derivation\end{tabular} & Max & Min & Definition & Source\\
\hline
\hline
\begin{tabular}{@{}c@{}}Mobile download \\ speed\end{tabular} & 26.81 & 14.80 & 65.41 & 6.32 & Average mobile download speed in Mbps & Speedtest Global Index (May 2019)\\\hline
Telecom Towers & 0.57 & 0.66 & 5.56 & 0.03 & \begin{tabular}{@{}c@{}} Number of installed telecom towers  \\ per 1000 mobile subscribers \end{tabular} & \begin{tabular}{@{}c@{}}Telecom Network Report Ed 4 (2017),\\ StatPlan Energy Research\end{tabular}\\\hline
Bandwidth & 191.11 & 131.29 & 685.00 & 25.00 & \begin{tabular}{@{}c@{}} Bandwidth for downlink cellular  \\ transmissions in MHz \end{tabular} & \begin{tabular}{@{}c@{}}National Frequency Allocation Table \\ (2018)\end{tabular}\\\hline
Telecom Papers & 3.84 & 3.17 & 10.49 & 0.00 & \begin{tabular}{@{}c@{}} Logarithm of the number of telecom \\ papers since 1948\end{tabular} & Web of Science\\\hline
\begin{tabular}{@{}c@{}}Telecom Papers \\ (Relative)\end{tabular} & -7.59 & 1.97 & -3.99 & -12.16 & \begin{tabular}{@{}c@{}} Logarithm of the ratio of Telecom Papers\\ to GDP (million USD)\end{tabular} & Web of Science\\\hline
Telecom TopPapers & 0.80 & 1.36 & 6.91 & 0.00 & \begin{tabular}{@{}c@{}} Logarithm of the number of telecom \\ papers cited over 100 times since 1948\end{tabular} & Web of Science\\\hline
\begin{tabular}{@{}c@{}}Telecom TopPapers \\ (Relative)\end{tabular} & -10.65 & 1.27 & -7.20 & -13.96 & \begin{tabular}{@{}c@{}} Logarithm of the ratio of Telecom \\ TopPapers to GDP (million USD)\end{tabular} & Web of Science\\\hline
ICT Patents & 3.44 & 3.57 & 12.33 & 0.00 & \begin{tabular}{@{}c@{}} Logarithm of the number of applications \\ for ICT related patents\end{tabular} & The Global IT Report (2016)\\\hline
\begin{tabular}{@{}c@{}}ICT Patents \\ (Relative)\end{tabular} & -7.84 & 2.49 & -3.24 & -13.70 & \begin{tabular}{@{}c@{}} Logarithm of the ratio of ICT patents\\ to GDP (million USD)\end{tabular} & The Global IT Report (2016)\\\hline
 \begin{tabular}{@{}c@{}}Technology \\ Availability\end{tabular} & 4.78 & 0.92 & 6.60 & 3.00 & \begin{tabular}{@{}c@{}} Rating of availability level of\\  latest technologies (from 1 to 7)\end{tabular} & The Global IT Report (2016)\\
\hline
\end{tabular}

\justify \textbf{Note}: 1) The data on mobile subscribers and GDP were obtained from ITU WTID (2018); 2) to maintain tractability and allow consistent analysis over all countries, countries reported having no research output are assigned with the value of one, so that the logarithm of the corresponding value equals zero; 3) the numbers of country-level ICT patent applications in 2012 and 2013 are available in the Global IT Report (2016), but historical data cannot be found. We thereby assume that the cumulative sum of ICT patent applications is ten times of that in 2012-2013; 4) for simplicity, the downlink bandwidth of time division duplexing (TDD) is assumed to be half of the full bandwidth. 
\end{table*}

The empirical model given in (\ref{ajhajhj2dsa2d1a}) is a reduced-form linear regression designed to assess the determinants of cross-national differences in the performance of cellular communications. We collect the latest datasets to construct a comprehensive country-level database, which covers 138 nations or regions (observations) and over 100 performance metrics in 2018. Metrics that lead to justifiable and robust empirical regularities are selected as key metrics. For clarity, the definitions and sources of the key metrics are listed in Table \ref{biaogeyi}.

We pay careful attention to the endogeneity and heterogeneity considerations, which are the main challenges confronting econometric applications. To avoid introducing multicollinearity, the estimations examine the proxies for research output separately and incorporate explanatory metrics that have independent implications. Table \ref{biaogeer} displays the estimates from the standard ordinary least squares (OLS) method, in which the first column is associated with the benchmark estimation, and other columns are related to estimations using alternative sets of metrics.

\begin{table}[!t]
\renewcommand{\arraystretch}{1.3}
\caption{Determinants of mobile download speed under the linear specification.}
\label{biaogeer}
\centering
\begin{tabular}{c|c|c|c|c}
\hline
\multirow{2}{*}{Metrics} & \multicolumn{4}{c}{Mobile download speed}\\
\cline{2-5}
 & 1 & 2 & 3 & 4\\
\hline\hline
\begin{tabular}{@{}c@{}} Telecom\\Towers \end{tabular} & \begin{tabular}{@{}c@{}} 4.140*** \\ (1.26) \end{tabular} & \begin{tabular}{@{}c@{}} 4.365***\\ (1.20)\end{tabular} & \begin{tabular}{@{}c@{}}4.746*** \\ (1.20)\end{tabular} & \begin{tabular}{@{}c@{}} 4.659***\\ (1.06)\end{tabular}\\\hline
Bandwidth & \begin{tabular}{@{}c@{}}0.0635*** \\ (0.01)\end{tabular} & \begin{tabular}{@{}c@{}} 0.0434***\\ (0.01)\end{tabular} & \begin{tabular}{@{}c@{}} 0.0402***\\ (0.01)\end{tabular} & \begin{tabular}{@{}c@{}} 0.0318**\\ (0.01)\end{tabular}\\\hline
\begin{tabular}{@{}c@{}} Telecom\\ Papers\end{tabular} & -- & \begin{tabular}{@{}c@{}}1.248*** \\(0.41) \end{tabular}& -- & -- \\ \hline
\begin{tabular}{@{}c@{}} Telecom\\ TopPapers\end{tabular} & -- & -- & \begin{tabular}{@{}c@{}}3.446*** \\(1.25) \end{tabular} & -- \\ \hline
ICT Patents & -- & -- & -- & \begin{tabular}{@{}c@{}}1.674*** \\(0.42) \end{tabular} \\ \hline
Constant & \begin{tabular}{@{}c@{}} 12.32***\\ (1.62)\end{tabular} & \begin{tabular}{@{}c@{}} 11.23***\\ (1.47)\end{tabular} & \begin{tabular}{@{}c@{}} 13.68***\\(1.68) \end{tabular} & \begin{tabular}{@{}c@{}} 12.16***\\(1.52) \end{tabular} \\ \hline
\begin{tabular}{@{}c@{}} Number of\\observations \end{tabular}& 138& 138& 138& 138\\\hline
R-squared & 0.412 & 0.453 & 0.472 & 0.501\\
\hline
\end{tabular}

\justify \textbf{Note}: 1) Robust standard errors in parentheses; 2) ***: $p<$0.01; **: $p<$0.05; *: $p<$0.1 ($p$ is the $p$-value in this statistical hypothesis).
\end{table}

According to the first estimation, mobile download speed is positively and significantly correlated with the number of installed telecommunication towers: each additional telecommunication tower per 1000 subscribers increases the download speed by 4.14 Mbps, implying a pronounced effect of telecommunication infrastructure on cellular network performance. Similarly, each 1 MHz increase in bandwidth leads to a 0.06 Mbps increase in the download speed. Overall, BS density and spectral resource can explain 41.2\% of the cross-national differences in mobile download speed, which validates the explanatory power (represented by the R-squared value in the table) of the two-factor model but also indicates the existence of other performance determinants. 

Based on the first estimation, estimations 2-4 further control for the cross-national differences in the PHY research outputs using different proxies. Overall, estimations using different proxies have reached robust results. As expected, the contribution of highly cited papers is larger than that of less cited papers, although the absolute quantity of highly cited papers is much smaller. The analytical results indicate that the PHY research has  statistically significant but only modest effects on the cellular network performance. In particular, the introduction of research outputs leads to only a 4-9\% increase in the explanatory power. Quantitatively, the estimated coefficient  suggests that a 1\% increase in the cumulative research outputs contributes to no more than a 0.04 Mbps increase in download speed. In other words, doubling the PHY research outputs will only lead to less than 4 Mbps increase in download speed (note that the estimated coefficient represents the marginal effect). This evidence lends credits to the speculation from Cooper's Law and the widespread concern over the increasing divergence between academia and industry \cite{4623708,5741160}. 

To approach the form of (\ref{ajhajhj2dsa2d1a}) and separate the effects of the BS density and spectral resource, we carry out the logarithmic transformation of (\ref{ajhajhj2dsa2d1a}) and obtain
\begin{equation}\small
\begin{cases}
&\log(C_i)=\beta_1\log\left(\frac{M_i}{U_iN_i}\right)+\beta_2\log(K_i)+\varepsilon_i\\
&\\
&\log(\mathrm{Speed}_i)=\beta_1\log(\mathrm{BaseStation}_i)\\
&+\beta_2\log(\mathrm{Spectrum}_i)+\beta_3\log(\mathrm{Research}_i)+\delta_i
\end{cases}
\end{equation}
Based on the log-log form, we re-estimate the parameters using the same econometric techniques. The re-estimation examines whether the results are sensitive to the selection of model specifications. More importantly, the re-estimation helps to quantify the performance gains of different inputs and thus makes the estimated coefficients more comparable. In the log-log specification, parameter $\beta_1$ represents the elasticity of download speed with respect to the number of BSs, which corresponds to the densification gain and so is the case of bandwidth and the PHY research output. Theoretically, $\beta_1$ and $\beta_2$ should approach unity, since the performance gains are presumably proportional as suggested by (\ref{djashk23ad111}). However, empirically, $\beta_1$ and $\beta_2$ can be less than one due to practical imperfections, but quantitative evidence remains scarce.

Table \ref{biaogesan} lists the estimated results under the log-log specification. Overall, the metrics have retained their explanatory power and statistical significance as in the former estimations. Quantitatively, a 1\% increase in the BS density leads to a 0.11-0.13\% increase in mobile download speed; the densification gain of spectral resource is relatively high (0.23-0.38\%) but also below the theoretical upper bound of unity. On the other hand, again, the contribution of the PHY research is estimated to be statistically significant but only marginal. A 1\% increase in the PHY research output is associated with an increase of no more than 0.11\% in cellular network performance. These findings indicate the possible existence of network resource missallocation, mismatch between research needs and research priority, and large potential for cellular performance improvement via research efforts.

\begin{table}[!t]
\renewcommand{\arraystretch}{1.3}
\caption{Determinants of mobile download speed under the log-log specification.}
\label{biaogesan}
\centering
\begin{tabular}{c|c|c|c|c}
\hline
\multirow{2}{*}{Metrics} & \multicolumn{4}{c}{Mobile download speed}\\
\cline{2-5}
 & 1 & 2 & 3 & 4\\
\hline\hline
\begin{tabular}{@{}c@{}} Telecom\\Towers \end{tabular} & \begin{tabular}{@{}c@{}} 0.133*** \\ (0.05) \end{tabular} & \begin{tabular}{@{}c@{}} 0.124***\\ (0.05)\end{tabular} & \begin{tabular}{@{}c@{}}0.121** \\ (0.05)\end{tabular} & \begin{tabular}{@{}c@{}} 0.110**\\ (0.05)\end{tabular}\\\hline
Bandwidth & \begin{tabular}{@{}c@{}}0.380*** \\ (0.06)\end{tabular} & \begin{tabular}{@{}c@{}} 0.297***\\ (0.07)\end{tabular} & \begin{tabular}{@{}c@{}} 0.291***\\ (0.06)\end{tabular} & \begin{tabular}{@{}c@{}} 0.225***\\ (0.06)\end{tabular}\\\hline
\begin{tabular}{@{}c@{}} Telecom\\ Papers\end{tabular} & -- & \begin{tabular}{@{}c@{}}0.0354** \\(0.02) \end{tabular}& -- & -- \\ \hline
\begin{tabular}{@{}c@{}} Telecom\\ TopPapers\end{tabular} & -- & -- & \begin{tabular}{@{}c@{}}0.107*** \\(0.03) \end{tabular} & -- \\ \hline
ICT Patents & -- & -- & -- & \begin{tabular}{@{}c@{}}0.0632*** \\(0.01) \end{tabular} \\ \hline
Constant & \begin{tabular}{@{}c@{}} 1.368***\\ (0.31)\end{tabular} & \begin{tabular}{@{}c@{}} 1.636***\\ (0.35)\end{tabular} & \begin{tabular}{@{}c@{}} 1.710***\\(0.33) \end{tabular} & \begin{tabular}{@{}c@{}} 1.889***\\(0.31) \end{tabular} \\ \hline
\begin{tabular}{@{}c@{}} Number of\\observations \end{tabular}& 138& 138& 138& 138\\\hline
R-squared & 0.424 & 0.446 & 0.469 & 0.520\\
\hline
\end{tabular}

\justify \textbf{Note}: 1) Robust standard errors in parentheses; 2) ***: $p<$0.01; **: $p<$0.05; *: $p<$0.1 ($p$ is the $p$-value in this statistical hypothesis).
\end{table}

\subsection{Robustness Tests}
Although insightful, the preliminary results in Table \ref{biaogeer} and Table \ref{biaogesan} could be misleading due to model misspecifications. Perhaps the most central concern is the existence of knowledge diffusion. That is, countries can benefit from technology innovations both within and across borders. Typical cases are the small-size developed countries, such as Luxembourg and the Netherlands, which have advanced mobile networks but rely little on independent research and development. Estimations that overlook cross-national technology absorption will overestimate the statistical significance and contribution of local research outputs. 

Another consideration is the issue of spurious correlation. That is, there might be positive correlations between research output, national productivity, and mobile network performance in the statistical sense, although the PHY research does not necessarily contribute to performance improvement. To put it simply, research output can be a redundant by-product in the production process, as indicated by the speculation over the death of the PHY research. Potential candidates are the largest economies, such as the U.S. and Japan, which have superior mobile networks, overcapacity, and substantial research outputs. The above considerations also cast doubts on the validity of cross-national comparisons among absolute levels of research output, since different countries differ considerably in economic scales. 

To test these hypotheses, the estimation incorporates the proxy called the availability level of latest technologies to control for the technology absorptive capacity (TAC) of each country, in contrast to countries' independent research outputs. Moreover, we construct relative research output as the ratio of research output to gross domestic product (GDP), which measures the strength of research efforts in each country. The transformation from quantity into quality metrics helps to control for economic scales and enable appropriate cross-national comparisons. Table \ref{biaogesi} displays the results of the misspecification tests, in which estimations 1-3 examine the potential overestimation of local research contribution (due to technology absorption) and estimations 4-6 further control for the scale economy effect using relative research outputs.

\begin{table*}[!t]
\renewcommand{\arraystretch}{1.3}
\caption{Results of model misspecification tests.}
\label{biaogesi}
\centering
\begin{tabular}{c|c|c|c|c|c|c}
\hline
\multirow{3}{*}{Metrics} & \multicolumn{6}{c}{Mobile download speed}\\
 &\multicolumn{3}{c}{\textit{Absolute research output}}&\multicolumn{3}{c}{\textit{Relative research output}}\\
\cmidrule(r){2-4}\cmidrule(l){5-7}
 & 1 & 2 & 3 & 4 & 5 & 6 \\
\hline\hline
\begin{tabular}{@{}c@{}} Telecom\\Towers \end{tabular} & \begin{tabular}{@{}c@{}} 4.379***\\ (1.32)\end{tabular} & \begin{tabular}{@{}c@{}} 4.559***\\(1.33) \end{tabular}& \begin{tabular}{@{}c@{}} 4.448***\\ (1.33)\end{tabular}& \begin{tabular}{@{}c@{}} 3.522***\\(1.32) \end{tabular}& \begin{tabular}{@{}c@{}} 3.689***\\ (1.36)\end{tabular}& \begin{tabular}{@{}c@{}} 3.847***\\(1.31) \end{tabular}\\\hline
Bandwidth & \begin{tabular}{@{}c@{}} 0.0179*\\(0.01) \end{tabular}& \begin{tabular}{@{}c@{}} 0.0191**\\(0.01) \end{tabular}& \begin{tabular}{@{}c@{}} 0.0201**\\(0.01) \end{tabular}& \begin{tabular}{@{}c@{}} 0.0163*\\(0.01) \end{tabular}& \begin{tabular}{@{}c@{}} 0.0298***\\(0.01) \end{tabular}& \begin{tabular}{@{}c@{}} 0.0199**\\(0.01) \end{tabular}\\\hline
\begin{tabular}{@{}c@{}} Telecom\\ Papers\end{tabular}  & \begin{tabular}{@{}c@{}} 0.742**\\ (0.35)\end{tabular}& --& --& \begin{tabular}{@{}c@{}} 1.726***\\(0.51) \end{tabular}& --&-- \\\hline
\begin{tabular}{@{}c@{}} Telecom\\ TopPapers\end{tabular}  & --& \begin{tabular}{@{}c@{}} 1.809**\\(0.83) \end{tabular}&-- &-- & \begin{tabular}{@{}c@{}} 1.427**\\(0.68) \end{tabular}&-- \\\hline
ICT Patents  &-- &-- &\begin{tabular}{@{}c@{}} 0.689*\\(0.37) \end{tabular} & --&-- & \begin{tabular}{@{}c@{}} 1.606***\\(0.50) \end{tabular}\\\hline
\begin{tabular}{@{}c@{}} Technology\\ Availability\end{tabular}  & \begin{tabular}{@{}c@{}} 7.650***\\(1.17) \end{tabular}& \begin{tabular}{@{}c@{}} 7.337***\\(1.20) \end{tabular}& \begin{tabular}{@{}c@{}} 6.905***\\(1.33) \end{tabular}& \begin{tabular}{@{}c@{}} 7.566***\\(1.14) \end{tabular}& \begin{tabular}{@{}c@{}} 7.986***\\(1.15) \end{tabular}& \begin{tabular}{@{}c@{}} 5.679***\\(1.35) \end{tabular}\\\hline
Constant & \begin{tabular}{@{}c@{}} -18.57***\\(4.80) \end{tabular}& \begin{tabular}{@{}c@{}} -15.97***\\(5.10) \end{tabular}& \begin{tabular}{@{}c@{}} -15.04***\\(5.47) \end{tabular}& \begin{tabular}{@{}c@{}} -1.434\\(7.17) \end{tabular}& \begin{tabular}{@{}c@{}} -4.044\\(8.83) \end{tabular}& \begin{tabular}{@{}c@{}} 6.211\\(9.28) \end{tabular} \\\hline
\begin{tabular}{@{}c@{}} Number of\\observations \end{tabular} &138 &138 &138 &136 &136 &136 \\\hline
R-squared & 0.586&0.587 &0.584 &0.600 &0.580 &0.598 \\
\hline
\end{tabular}

\justify \textbf{Note}: 1) Robust standard errors in parentheses; 2) ***: $p<$0.01; **: $p<$0.05; *: $p<$0.1 ($p$ is the $p$-value in this statistical hypothesis); 3) for the estimations of relative research output, there are only 136 observations.
\end{table*}

Overall, mobile network performance relates closely to the TAC: a unit increase in the availability level of the latest technologies leads to a 5.68-7.65 Mbps increase in mobile download speed. As expected, the statistical significance of other determinants reduces as  technology availability enters the model, since these metrics simultaneously account for the development level of telecommunication technology, and thus generate multicollinearity. Due to the same reason, estimations 1-3 lead to lower estimates of local research contribution than the previous results. These findings suggest that the model has properly controlled for the effect of technology absorption.

Furthermore, according to estimations 4-6, alternative definitions of research output do not alter the main conclusions drawn by the previous data analysis. These estimations reject the hypothesis of spurious correlation and verify the presence of technology absorption as well as the rate of return on the domestic PHY research. Overall, the empirical analysis points towards robust empirical relationships between mobile download speed, telecommunication infrastructure, spectral resource, and the PHY research. It has been observed that the three-factor empirical model can explain over 50\% of the variations in cellular network performance.

\section{Suggestions for the Stakeholders of the PHY Academic Research}
Similar to the business world, in academia, the criterion of resource allocation is associated with the cost-benefit evaluation of research investment, without which the academic policy-making process can be subjective and counterproductive. On the other hand, the wireless communication community has long sensed the diminishing return of the PHY research \cite{4623708,5741160}. According to our estimates, the marginal return of the PHY research is highly statistically significant but quantitatively small: a 1\% increase in the cumulative research output leads to less than a 0.1\% performance improvement. Surprisingly, the marginal return of highly cited papers is about three times as that of less cited papers, although the number of highly cited papers is only one-twentieth as that of less cited papers and ICT patents. Accordingly, given that marginal contribution represents the effect of percentage increases in research output, each highly cited paper is equivalent to 60 less cited papers or 30 ICT patents in terms of the payoff of research investment. In this sense, the value of path-breaking research is self-evident.

Today, we need to confront the fact that the impressive performance improvement in the past decades is only marginally attributable to the PHY research. The top priority is not to blame someone for the stasis in the PHY research (indeed, the stasis can be an inevitable feature of disciplines that approach maturity), but to balance the resource allocation towards the sub-disciplines that hold great potentials. With this purpose, we provide a series of suggestions for the stakeholders of the PHY  research in the following subsections.

\subsection{Policy Makers}
There is no doubt that administrative intervention has a significant impact on applied disciplines, including wireless communications. Policy orientation should focus on not only the prospective wireless technologies that do not have solid foundation nowadays, but also on the practical wireless technologies that can benefit the society in the foreseeable future. Meanwhile, as network speed and throughput can be well satisfied by building more BSs and expanding spectrum (at lease for 5G and 6G) \cite{6736747}, the PHY research should be justified by the improvements on other performance evaluation metrics. Consequently, more feedback and needs from the end beneficiaries of the PHY research should be quantified and taken into account. Under such a need-oriented ideology, user's satisfaction of service (SoE) would be adopted as a pivotal metric for making policies and industrial plans. For example, in addition to the conventional metrics (e.g., network speed, throughput, and latency), security, secrecy, and privacy can be quantified and adopted as the key features when blueprinting next generation networks. In addition, governmental subsidies should be provided to lead indirectly to global profits in accordance with the Sustainable Development Goals (SDGs) set by the United Nations General Assembly in the 2030 Agenda.

\subsection{Funders}
When assessing a project proposal applying for financial support, funders should pay attention to not only how many papers will be published or how many patents will be applied at the end of the project cycle, but also the possibility and vision of technology transfer as well as profitability. The assessment of potential market value and practicability should be based on the real needs of end users, which is an important reference point of funding applications. It is highly suggested to add these economic features as part of funding strategies and carry out a long-term tracking and evaluation of them. Then, a new evaluation system of the research outcomes could be constructed to appraise how good a project has been from both technological and socio-economic perspectives. Furthermore, funding priority should be given to the projects contributing to the SDGs. In summary, funding systems have to be efficient and instructive, since they serve as the core mechanism in stimulating creative thinking and innovative technologies in the ongoing 5G evolution.

\subsection{Researchers}
To some extent,  telecommunications are becoming a subject of esoteric mathematical models and far-fetched theories, which renders the field gradually divorced from reality. Many telecommunication theories are barely understandable to telecommunication practitioners, and their applicability has not been proved yet. Consequently, new research directions are the key to restore the discipline to its former vitality. The real problem is not why we have devoted so much to the PHY research, but why we have so far de-emphasized the non-PHY research and the needs from end beneficiaries. Accordingly, wireless research community needs to pay more attention to alternative approaches to the fundamental questions of telecommunication science. End beneficiaries should also have a voice in shaping research agendas. Researchers with different backgrounds should also be encouraged to address the same problems using multidisciplinary approaches, such as the empirical methods that can serve as a bridge between academic and real-world problems. Researchers might also borrow the PESTEL analysis (PESTEL: an acronym that stands for \textbf{P}olitical, \textbf{E}conomic, \textbf{S}ocial, \textbf{T}echnological, \textbf{E}nvironmental and \textbf{L}egal factors) from business studies to appraise the prospect of a wireless technology, instead of simply focusing on the academic scholarship and/or the profitability of mobile network operators.

\section{Closing Remarks and Future Work}\label{c}
In this article, we employed econometric methodologies to investigate the usefulness of the PHY academic research and compared it with other non-technological factors in terms of their contribution to the cellular network performance. The empirical analysis confirmed that the main contributor of the service improvement is the densification of cellular network, whereas only a marginal portion of the service improvement can be attributed to the PHY academic research. The empirical evidence also suggests that highly cited publications have much larger and more statistically significant impacts on cellular network performance than other research outputs. Overall, the analysis has documented robust empirical relations between network capacity, BS density, spectral resource, and the PHY research activities. Based on these findings, we provided a series of suggestions for the stakeholders of the PHY academic research.

Turn to the core issue under discussion: is the PHY dead? In conclusion, NO, the PHY research is not dead yet, as the marginal return of the PHY research remains statistically significant, and there is indeed huge potential for performance improvements via research efforts (as suggested by the low densification gains of telecommunication infrastructure and spectral resource). However, we also need to confront and react to the diminishing return of the PHY research. Close collaboration between policy makers, funders, and researchers is indispensable to restore the quondam vitality of the PHY research. Through the results and discussions presented in this article, it is reasonable to state that \textit{the PHY research is sick but curable}.

Employing empirical analysis based on actual observations, researchers are able to answer real-world problems such as the relative importance of academic research, other than being confined to the mathematical problems of interest only to academics. This study explores a new angle to quantify the marginal contributions of BS density, spectral resource, and research output in promoting telecommunication performance. This study is helpful to reconcile the controversy on the usefulness of the PHY research. 

However, it is important to stress that our analysis remains preliminary and tentative. The empirical analysis has focused on cellular communications, which left other branches of wireless communications unexplained (e.g., machine-to-machine (M2M) communications and satellite communications). In addition, the absence of reliable data has discouraged the research efforts into important topics, such as the impacts of intellectual property protection, technological standardization, and national innovation strategy on the cellular network performance. Also, due to the data insufficiency, our attempts are limited to simple empirical regularities, other than rigorous causal relations between research investment and performance improvement. All in all, as we are entering the era of `Big Data,’ the PHY research should center more on the exploitation of statistical systems, which will enrich the analytical tools to explore promising research directions hidden in telecommunication science.

\bibliographystyle{IEEEtran}
\bibliography{bib}

\begin{IEEEbiographynophoto}{Kevin Luo} (kevin.luo@stu.kobe-u.ac.jp) received his bachelor degree in Information Management from North China Electric Power University in 2014 and master degree in Economics from Kobe University in 2017. He is currently a PhD candidate in Economics with Kobe University. He also serves as teaching and research assistant at Kobe University for a number of graduate and undergraduate courses. He won the Kishimoto Award 2019 for his academic achievements. His research interests include empirical economics, behavioral economics, and econometrics.
\end{IEEEbiographynophoto}

\begin{IEEEbiographynophoto}{Shuping Dang} [M'18] (shuping.dang@kaust.edu.sa) received a B.Eng (Hons) in Electrical and Electronic Engineering from the University of Manchester (with first class honors) and a B.Eng in Electrical Engineering and Automation from Beijing Jiaotong University in 2014 via a joint `2+2' dual-degree program, and a D.Phil in Engineering Science from University of Oxford in 2018. He joined in the R\&D Center, Huanan Communication Co., Ltd. after graduating from University of Oxford, and is currently working as a Postdoctoral Fellow with the Computer, Electrical and Mathematical Science and Engineering (CEMSE) Division, King Abdullah University of Science and Technology (KAUST).
\end{IEEEbiographynophoto}
\begin{IEEEbiographynophoto}{Basem Shihada}[SM'12] (basem.shihada@kaust.edu.sa) is an associate and founding professor of computer science and electrical engineering in the Computer, Electrical and Mathematical Sciences and Engineering (CEMSE) Division at King Abdullah University of Science and Technology (KAUST). Before joining KAUST in 2009, he was a visiting faculty at the Computer Science Department in Stanford University. His current research covers a range of topics in energy and resource allocation in wired and wireless communication networks, including wireless mesh, wireless sensor, multimedia, and optical networks. He is also interested in SDNs, IoT, and cloud computing. In 2012, he was elevated to the rank of Senior Member of IEEE.
\end{IEEEbiographynophoto}
\begin{IEEEbiographynophoto}{Mohamed-Slim Alouini}[F'09] (slim.alouini@kaust.edu.sa) received the Ph.D. degree in electrical engineering from the California Institute of Technology (Caltech), Pasadena, CA, USA, in 1998. He served as a faculty member at the University of Minnesota, Minneapolis, MN, USA, then at Texas A\&M University at Qatar, Education City, Doha, Qatar, before joining King Abdullah University of Science and Technology (KAUST), Thuwal, Makkah Province, Saudi Arabia as a professor of electrical engineering in 2009. At KAUST, he leads the Communication Theory Lab and his current research interests include the modeling, design, and performance analysis of wireless communication systems.
\end{IEEEbiographynophoto}

\end{document}